\documentclass[aps,prl,twocolumn,superscriptaddress]{revtex4-1}
\usepackage[utf8]{inputenc}  
\usepackage[T1]{fontenc} 
\usepackage{amsmath}
\usepackage{amssymb}
\usepackage{amsfonts}
\usepackage{amsbsy}
\usepackage{graphicx}
\usepackage{color}

\newcommand*{\rom}[1]{\expandafter\@slowromancap\romannumeral #1@}
\DeclareMathOperator{\e}{e}
\begin{document}

\title{Observation of Extra Photon Recoil in a Distorted Optical Field}

\author{Satyanarayana Bade}
\affiliation{Laboratoire Kastler Brossel, Sorbonne Universit\'e, CNRS, ENS-PSL University, Coll\`ege de France, 4 place Jussieu, 75005 Paris}
\author{Lionel Djadaojee}
\affiliation{Laboratoire Kastler Brossel, Sorbonne Universit\'e, CNRS, ENS-PSL University, Coll\`ege de France, 4 place Jussieu, 75005 Paris}
\author{Manuel Andia}
\altaffiliation{Current address: University of Strathclyde, Department of Physics, SUPA, Glasgow G4 0NG, United Kingdom}
\affiliation{Laboratoire Kastler Brossel, Sorbonne Universit\'e, CNRS, ENS-PSL University, Coll\`ege de France, 4 place Jussieu, 75005 Paris}
\author{Pierre Clad\'e}
\email{pierre.clade@lkb.upmc.fr}
\affiliation{Laboratoire Kastler Brossel, Sorbonne Universit\'e, CNRS, ENS-PSL University, Coll\`ege de France, 4 place Jussieu, 75005 Paris}
\author{Sa\"{\i}da Guellati-Khelifa}
\email{saida.guellati@lkb.upmc.fr}
\affiliation{Laboratoire Kastler Brossel, Sorbonne Universit\'e, CNRS, ENS-PSL University, Coll\`ege de France, 4 place Jussieu, 75005 Paris}
\affiliation{Conservatoire National des Arts et M\'etiers, 292 rue Saint Martin, 75003 Paris, France}

\date{\today}

\def\kperp{\ensuremath{\mathbf{k_\perp}}}
\def\abskperp{\ensuremath{||\kperp||}}

\begin{abstract}
Light carries momentum which induces on atoms a recoil for each photon absorbed. In vacuum, for a monochromatic beam of frequency $\nu$, the global momentum per photon is bounded by general principles and is smaller than $h \nu/c$ leading to a limit on the recoil. However, locally this limit can be broken. In this paper, we give a general formula to calculate the recoil in vacuum. We show that in a laser beam with a distorted optical field, there are regions where the recoil can be higher than this limit. Using  atoms placed in those regions we are able to measure directly the extra recoil. 
\end{abstract}


\maketitle

Light exerts upon a surface a pressure called the radiation pressure. From a classical perspective, this pressure is described by a linear momentum with a density that is proportional to the energy flux, the Poynting vector \cite{jacksonclassical1999}. From a quantum mechanical point of view, the pressure corresponds to momentum carried by photons. The momentum $\vec{p}$ is defined only for a plane wave and has an amplitude proportional to the energy (or the frequency $\nu$) of the photon $||\vec{p}|| = h\nu/c$. 

When an atom of mass $m$ absorbs a photon, the momentum of the photon induces a recoil  on the atom. The recoil velocity, $v_r = h\nu/mc$, plays an important role in atomic physics (laser cooling and atom interferometry). The recoil velocity has been measured with a very high accuracy for the determination of the fine structure constant \cite{Bouchendira2011, Wicht:02, Muller2010, Parker191}. Measuring the recoil of atoms is also a tool to probe the local momentum of light. Even though the absorption of photons is quantized, the best way to calculate the recoil velocity is to use the classical momentum given by the ratio between the density of linear momentum and the photon density.  This approach reveals interesting properties of light,  such as in the evanescent regime where a recoil velocity larger than $h\nu/mc$ (i.e an extra recoil velocity) is expected \cite{HUARD1978, MATSUDO199864}. 

In this paper, we show that the extra recoil is actually not limited to the case of an evanescent wave: even in a beam propagating in vacuum the momentum of a photon can be locally  higher than $h\nu/c$. Using atoms placed in these locations, we have been able to confirm experimentally this effect.

For a linearly polarized beam and under the paraxial approximation one can use the scalar diffraction theory to describe the optical field \cite{BornWolf:1999:Book}. The momentum of a photon is given by the canonical momentum $\vec{p}=\hbar \vec{\nabla }\phi$, where $\phi$ is the spatial phase of the laser \cite{Antognozzi2016}\footnote{More precisely the phase gradient is called the canonical momentum, to obtain the total momentum we need to add spin momentum. In this article we neglect this term.}. Let us consider a laser beam propagating along the $z$ axis of which we know the amplitude $A(x, y, z_0)$ and phase $\phi(x, y, z_0)$ on the plane $z=z_0$. One can use the Helmholtz equation to propagate the wavefront along the $z$ axis and  therefore deduce the $z$ component $p_z$ of the momentum at position $z_0$, $p_z=\hbar k (1+\delta k_\mathrm{rel})$ with 
\begin{equation}
\label{eq:momentum_correction}
\delta k_{\mathrm{rel}} = -  \frac12\left|\left|\frac{\vec\nabla_{\perp}\phi}{k}\right|\right|^2+\frac1{2k^2}\frac{\Delta_{\perp} A}{A}
\end{equation}
where $\Delta_{\perp}$ and $\vec{\nabla}_{\perp}$ are evaluated in the plane $z=z_0$ and $k=2 \pi \nu/c$. The first term related to the phase gradient corresponds to a tilt in the propagation direction with respect to the $z$ axis due to a local distortion of the wavefront. The second term gives a correction to the momentum even in the case of a plane wavefront. This counter intuitive term is actually a generalization of effects already identified in situations where an analytical solution exists, such as evanescent and Gaussian beams.

In the case of an evanescent wave $A_0 e^{\alpha y - i\beta z}$, the longitudinal momentum, proportional to $\beta$, is usually obtained with the dispersion relation ($\beta^2 - \alpha^2 = k^2$). At first order in $\alpha$ (paraxial approximation), this equation leads to $\beta/k  = 1+ \alpha^2/2k^2$, the value obtained by calculating the Laplacian of the amplitude. In \cite{MATSUDO199864}, the extra recoil observed in the saturation spectroscopy was explained by the pseudo-momentum of the photon in the refractive medium used to generate the evanescent wave. We should notice that the extra recoil can be explained by Eq.~(\ref{eq:momentum_correction}) which does not imply to be close to the surface of the refractive medium. As we will see, even in vacuum, positive correction to the recoil could be observed. 

For a Gaussian beam, at the position of the waist, $\vec\nabla_{\perp}\phi = 0$, and the momentum of light is given by the Laplacian of the amplitude. This leads to a correction of the momentum  
\begin{equation}\label{eq:gouy}
\frac{p_z}{\hbar k} = 1 - \frac 2{k^2}\left(\frac1{w^2}-\frac{r^2}{w^4}\right)
\end{equation}
which gives a correction to the atomic recoil, the well known Gouy phase correction \cite{Wicht:02, PhysRevA.72.023602, clade:052109}, calculated using the analytical value of the phase of a Gaussian beam. Atoms being placed usually at the center of a Gaussian beam experience a negative correction. This correction is interpreted with the dispersion in momentum of the plane wave decomposition of the laser beam \cite{Feng:01}. Each plane wave  tilted by an angle $\theta$ with respect to the $z$-axis gives a negative correction $-\theta^2/2$. The average longitudinal momentum reduction is $\delta k_\mathrm{rel} = -1/(kw)^2$. Locally, the correction is different: at the center the correction is twice this value and at large distances, the sign of the correction changes: even for a Gaussian beam, an extra recoil can be observed. 

In this paper we consider  a more general case of distorted optical field with amplitude noise and phase noise. We should first note that the spatial phase and amplitude fluctuations are mixed during propagation. Even a wave-front distortion caused by optics will appear as amplitude fluctuations after propagation. Furthermore both the relative fluctuations of the amplitude ($\sigma_A/A$) and the phase fluctuations ($\sigma_\phi$) have the same standard deviation. In Eq.~(\ref{eq:momentum_correction}), the term due to wave-front distortions (which scales as $\sigma_\phi^2$) is therefore negligible compared to the term due to amplitude fluctuations, which scales as $\sigma_A/A$. This leads to a counter intuitive result: the term due to spatial intensity fluctuations, which is usually ignored, is dominant over the term due to wave-front distortions. In the rest of the paper, the latter is always neglected. 

In order to estimate the correction due to the local fluctuations of the amplitude  we consider, for the sake of simplicity,  an optical wave with an auto-correlation function of the amplitude  given by $\sigma_i^2 e^{-(x^2 + y^2)^2/l^2}/4$. The parameter $l$ is the correlation length that  gives the scale of the second term of Eq.~(\ref{eq:momentum_correction}) and $\sigma_i$ is the RMS relative intensity noise. 
Using the wave-vector power spectral density (WV-PSD) we estimate $\sigma(\delta k_\mathrm{rel})$ at  $\frac{\sqrt{2}\sigma_i}{(kl)^2}$ (for details see \cite{Sup}). It is important to note that this term is different from the average momentum of the photons that can also be calculated from the WV-PSD, yielding $-\frac{\sigma_i^2}{2(kl)^2}$. This result is related to the previous one: even though the typical value of the longitudinal momentum of plane-wave photons scales as $\sigma_i^2/(kl)^2$, due to coherence between the different plane waves, the local momentum fluctuates with a much higher amplitude. 

\begin{figure*}
\includegraphics[width=.88\linewidth]{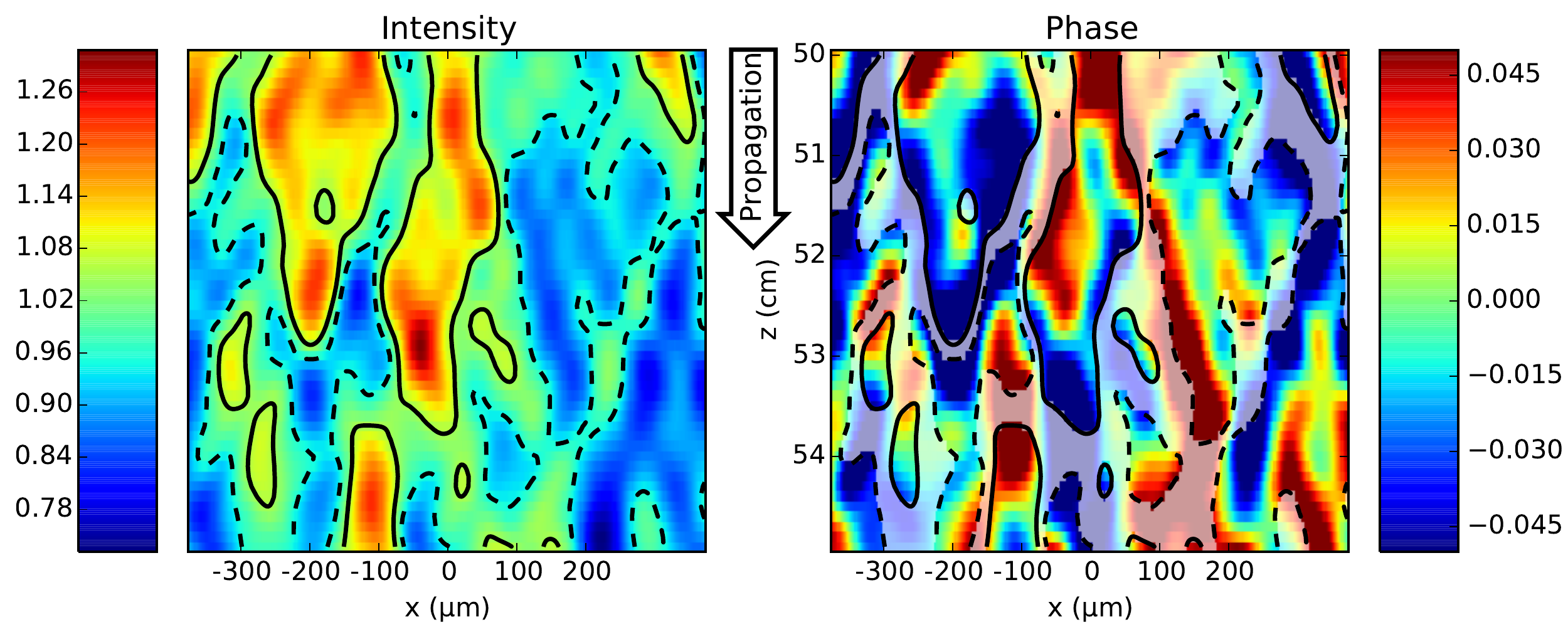}
\caption{\label{correlation_picture} Simulation of a slice of the intensity (left) and phase (right) of a laser beam taken along the propagation axis (vertical axis in the figure). Random phase fluctuations with a typical size of 100 $\mathrm\mu m$ were generated on the $(x,y)$ plane and propagated over 50~cm. The continuous (resp. dashed) contour shows regions of relatively high (resp. low) intensity. In the high intensity region (resp. low intensity), the phase evolves along the z direction, from a typical value of 50 mrad to -50 mrad (resp. -50~mrad to 50~mrad), leading to a local variation of the photon momentum.}
\label{correlation_picture}
\end{figure*}

Let us take for example $l=100~\mathrm{\mu m}$ and RMS amplitude of $\sigma_i=0.05$ for  a light wave-length of  $\lambda=780$ nm. This corresponds to typical values when using poor quality optical elements. The RMS value of $\delta k_{\mathrm{rel}}$ is 100~ppb. The amplitude of this effect is much larger than the sensitivity of atom recoil measurements performed using atom interferometry. However it has never been observed yet because experiments measure the average recoil and are not sensitive to local fluctuations.

To observe  this effect one can take advantage of the correlation between local light momentum and light-field intensity: indeed, this correlation arises from the Laplacian term in equation~\ref{eq:momentum_correction}, which changes sign between minima and maxima of intensity.  To illustrate and understand this correlation and how the experiment takes advantage of it, we have performed a numerical simulation of the propagation of a noisy light profile. We generated a field on the $(x,y)$ plane with a uniform intensity and a random phase noise with a typical correlation length $l=100~\mathrm{\mu m}$ and RMS amplitude of $0.07~\mathrm{rad}$. Figure~\ref{correlation_picture} shows a slice of this beam taken in the $(x, z)$ plane after a propagation of 50 cm. The high or low intensity regions have a typical width of $l=100~\mathrm{\mu m}$ and a typical length of $z_l= l^2/\lambda\simeq 1~\mathrm{cm}$. On the right of the picture, we have plotted the phase of the laser beam only for those regions. For high intensity regions (continuous contour), the phase evolves from typically 0.05~rad on the top to -0.05~rad at the bottom (red to blue, or $\sigma_i$ to $-\sigma_i$): there is a negative correction to the photon momentum with a relative amplitude that scales as $\sigma_i/ kz_l$, i.e. $\frac{\sigma_i}{(kl)^2}$ as calculated above. The opposite occurs for low intensity regions (dashed line).

Therefore, in order to see this effect, one has to select atoms in regions of relatively high or relatively low intensity.  For that let us  consider an experiment sensitive to the momentum of light: if the probability $P(I)$ to transfer light momentum to the atoms depends on the intensity, the spatial distribution of atoms is filtered and is therefore correlated to the local photon momentum. This gives rise to a correction with respect to the average momentum 
\begin{equation}
\label{eq:systematic_effect}
\langle \delta k_{\mathrm{rel}} \rangle =\frac{\langle \delta k_{\mathrm{rel}}  P(I) \rangle} {\langle  P(I) \rangle}
\end{equation}

As  $I$ and $\delta k_{\mathrm{rel}}$ are anti-correlated (see \cite{Sup}), the correction due to local fluctuations of the light momentum can be negative or positive depending on whether $P(I)$ is a decreasing or an increasing function of intensity. This is what we observed in the experiment described below.

\begin{figure}
\includegraphics[width=.88\linewidth]{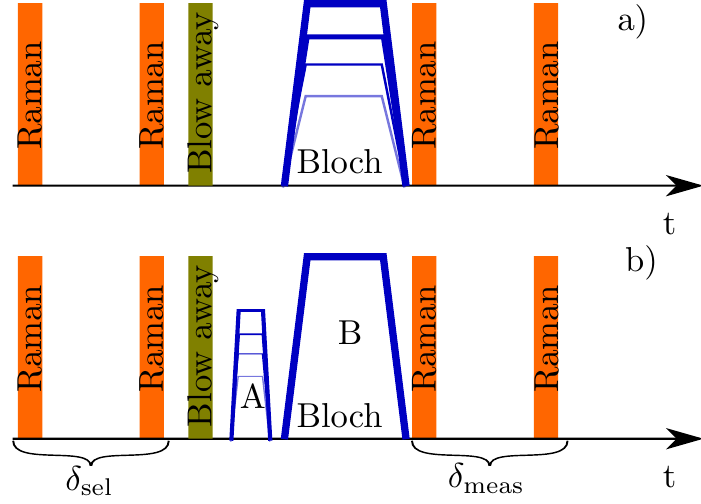}
\caption{\label{fig:sequencehsurm} Temporal sequence for the measurement of the recoil velocity. The Ramsey-Bord\'e interferometer consists of four Raman $\pi/2$ pulses. Between the first and second pair of pulses, BOs are used to accelerate atoms. In a),  the intensity of the Bloch beams is reduced in order to select atoms in the high intensity region. In b), the BO pulse A is used to remove atoms in the high intensity region and the pulse B at fixed intensity is used to the measurement of the recoil velocity.}
\end{figure}

Our experiment uses Bloch oscillations (BO) in an accelerated optical lattice \cite{BenDahan,Wilkinson} to transfer a large number of recoils to ultra-cold $^{87}\mathrm{Rb}$ atoms. The recoil velocity is then measured via the Doppler effect using atom interferometry. A detailed description of the experimental set-up can be found in \cite{Bouchendira2011}. We use an atomic cloud of nearly 10$^6$ atoms with a 2 mm radius and at a temperature of 4 $\mu$K. The interferometer is operated in the usual Ramsey-Bord\'e configuration with a Ramsey time of 10 ms. It uses two pairs of $\pi/2$ light pulses that drive a Raman transitions between the hyperfine levels $F=1$ and $F=2$. The first pair selects atoms which are then coherently accelerated with 250 BO in 3.5 ms, followed by the second pair that measures the total momentum transferred between the two pairs of pulses. Interference fringes are obtained by scanning the frequency $\delta_\mathrm{meas}$ of the last pair of pulses (see Fig.~\ref{fig:sequencehsurm}) and by measuring the relative population in each hyperfine level ($F=1$ and $F=2$) using time-of-flight technique.

The efficiency of BOs depends sharply on the intensity of the laser. If the intensity is lowered, due to Landau-Zener losses only the atoms in the higher intensity regions survive the BO and are involved in the interferometer. For instance, if 50\% of the atoms are lost during BO, the remaining atoms are mainly in the regions where the intensity is larger than the average and a negative effect of the order of $\frac{\sigma_i}{(kl)^2}$ should be observed.

\begin{figure}
\includegraphics[width=\linewidth]{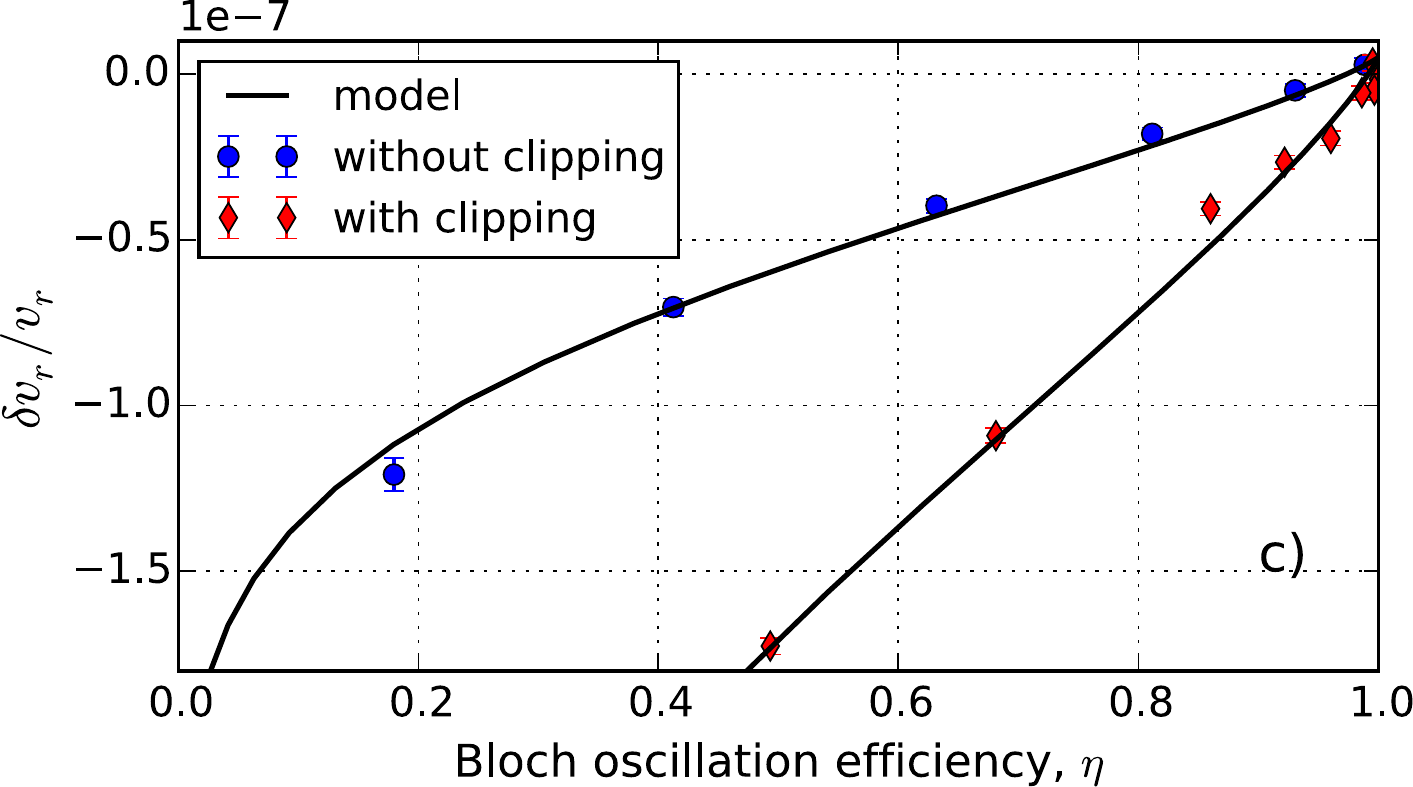}
\caption{\label{usual_effect_fit}  The relative recoil velocity variation $\delta v_r/v_r$ as a function of  efficiency $\eta$.  Dots correspond to experimental data which are obtained  in two cases: with and without clipping one of the Bloch beams. The deviation in the momentum from the ideal plane wavefront is larger when the beam is clipped due to amplified fluctuations.  Line: calculated correction due to the change in momentum $\delta k_{\mathrm{rel}}$ as a function of the efficiency of BOs (from Eq.~\ref{eq:effect_i}). In our model, this curve is universal. The only parameters are the RMS intensity fluctuations $\sigma_i$ and the correlation length $l$ of those fluctuations which give the amplitude factor ($\sigma_i/(kl)^2$) of the correction.}
\end{figure}

We repeat the recoil measurement by scanning the intensity of the Bloch beams (see Fig.~\ref{fig:sequencehsurm}.a), we measure also the fraction of atoms that survive BOs, this fraction gives the efficiency $\eta$ of BOs. Dots in Fig.~\ref{usual_effect_fit} show the relative recoil velocity with respect of  the value obtained in \cite{Bouchendira2011} as function of $\eta$. We choose this representation because, as we will see latter,  in our model this curve is universal.  As the intensity is lowered, the BOs efficiency decreases (see \cite{Sup}) and  the deviation becomes larger. In order to confirm the effect, one of the two Bloch beams is clipped with an aperture of diameter 3 mm to amplify the intensity fluctuations. Here, we clearly observe an increased deviation from the plane wave momentum when the beam is clipped for a given $\eta$.

In order to observe an extra recoil, one has to place atoms in regions of relatively low intensity. To this end, we have changed the temporal sequence of the experiment (Fig.~\ref{fig:sequencehsurm}.b): just before the acceleration, we perform 10 BOs with a tunable intensity (Bloch pulse-A) in such a way that atoms performing BOs are removed (pushed away, they do not participate in the subsequent recoil measurement). At low intensities, no atoms are removed and hence no effect should be observed.  When the power of this first Bloch beam is increased, atoms are removed from the experiment, starting with the ones that are at maxima of intensity. For higher power, only atoms at minima survive. The first Bloch pulse, used to filter the spatial distribution of atoms, is then followed by the 250 BOs at a fixed and high power which are normally used to measure precisely the recoil velocity  (Bloch  pulse-B on Fig.~\ref{fig:sequencehsurm}.b).

\begin{figure}
\includegraphics[width=.88\linewidth]{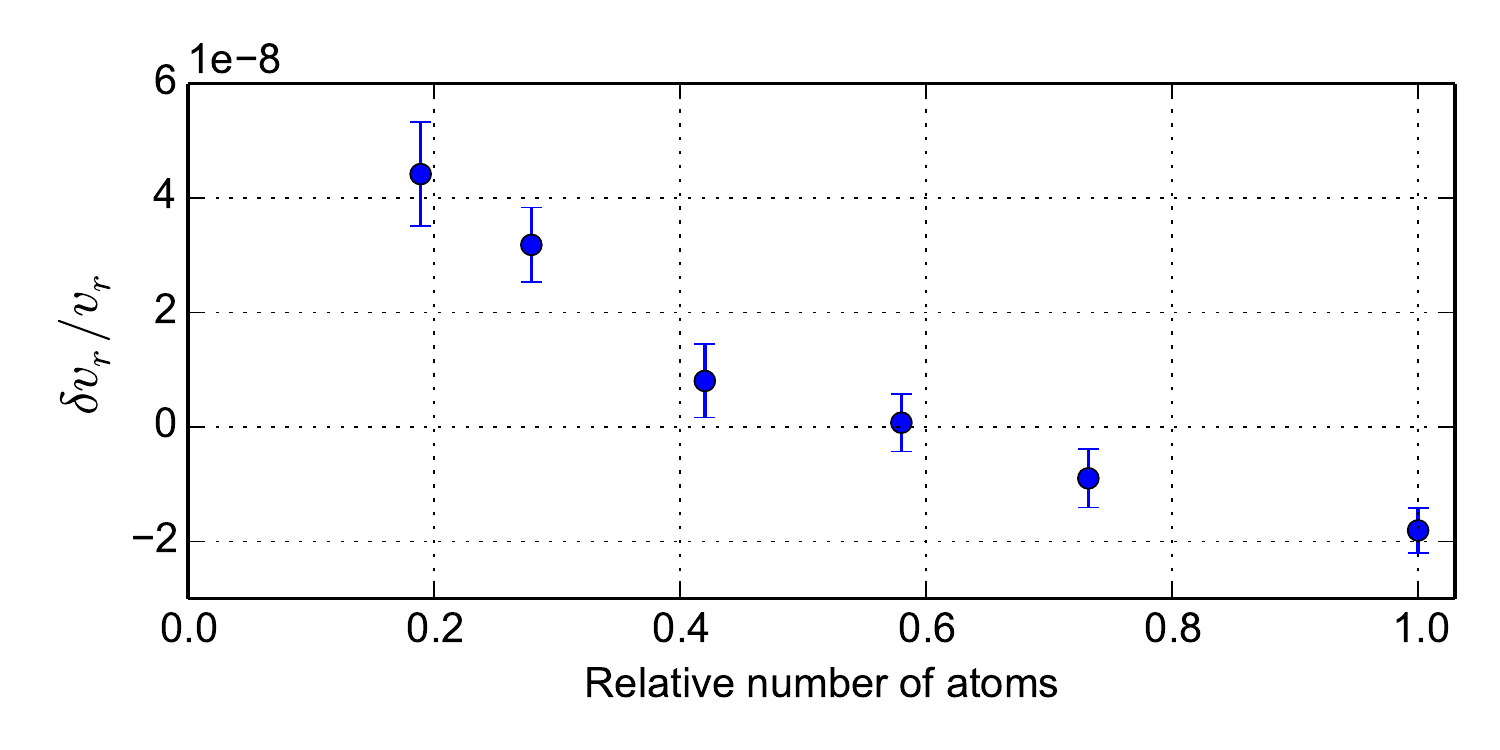}
\caption{\label{positive_effect} Measurement of the relative recoil in a distorted field obtained by removing atoms in the high intensity regions of the Bloch beams, following the protocol described on Fig. \ref{fig:sequencehsurm}.b.}
\end{figure}

Figure~\ref{positive_effect} shows the results. This experiment was performed with one of the beams being clipped by a 3~mm aperture. Without losses, we observe a momentum slightly smaller than $\hbar k$, this is the same effect described earlier. However, when the number of atoms is reduced by the selecting BOs, the measured photon momentum increases, leading to an effect larger than $4\times 10^{-8}$. One would expect to see a correction with an opposite sign and the same amplitude in comparison with the data shown in Fig.~\ref{usual_effect_fit}, but we observe a reduction of the amplitude of the effect that can be explained by the motion of atoms between the selection and measurement pulses: even if they are selected at low intensity, the transverse motion of the atoms brings them to regions of higher intensity, leading to a reduction of the correction. 

Let us now discuss quantitatively the experimental data. For BOs, the probability $P(I)$ is governed by the Landau-Zener losses. As it varies sharply with the intensity, we model it with a Heaviside function centered around a critical intensity $I_c$ \cite{PhysRevA.95.063604, Peik}. Using this model, one can calculate an analytical expression for the efficiency $\eta$ and average  $\delta k_{\mathrm{rel}}$ (from Eq.\ref{eq:systematic_effect}) as a function of $I_{\mathrm{rel}} = (I_c - I)/(I \sigma_i)$ 

\begin{equation}
\label{eq:eta_i}
\eta = \frac{\mathrm{erfc}(I_{\mathrm{rel}}/\sqrt{2})}{2}
\end{equation}

\begin{equation}
\label{eq:effect_i}
\langle \delta k_{\mathrm{rel}} \rangle = -\frac{1}{k^2 l^2 } \frac{\sigma_i} {\sqrt{\pi/2}} \frac{\e^{-I^2_{\mathrm{rel}}/2}}{\eta}
\end{equation}
From these two quantities a parametric curve can be drawn and is used to fit the data in Fig.~\ref{usual_effect_fit}. A single parameter is used, the quantity $\frac{\sigma_i}{(k l)^2 }$. For the data without beam clipping, we obtain $\frac{\sigma_i}{(k l)^2 }=8\times 10^{-8}$, this corresponds to $5\%$ of fluctuations at a scale of $l=100~\mathrm{\mu m}$, which is compatible with an independent measurement of the phase profile.

Since BOs are a pure coherent  quantum phenomenon, one can think that the effect described and observed in this paper  is rather a signature of processes which cause a decoherence of the system (spontaneous emission, inter-atomic interactions or disordered light potential). Since the first observation of BOs of cold atoms in optical lattices an important work has been devoted to investigate the question of  decoherence   \cite{Kolvsky2002, KOLOVSKY2004, Roati2004, Lye2007, Schulte2006, Schulte2008, DampBO2008, ReneeChar2012}. In particular, the dynamics of BOs in disordered optical lattices has been investigated in detail in \cite{Schulte2008}. Both theory and experiment \cite{Schulte2006, Schulte2008, DampBO2008} have shown that even a very small disorder results in broadening of the quasi-momentum distribution that causes a damping of the center-of-mass motion. In our experiment during the Bloch acceleration atoms move by about 5 mm, this distance is smaller than the characteristic length of the intensity fluctuations along the $z$-axis (see figure \ref{correlation_picture}).  In addition, we analyzed in detail our raw experimental data (time-of-flight signals and the atomic fringes), as shown in \cite{Sup} we do not observe any signature of the scattering of the quasi-momentum underlying  a damping of  BOs. Furthermore, we used a dilute atomic gas and the Bloch laser frequencyl is far-detuned from the resonance, thus inter-atomic interactions and spontaneous emission are  negligible \cite{Bouchendira2011}.

In this paper, we have calculated and measured locally the momentum of light using the recoil of atoms. We have been able to measure a recoil higher than the fundamental limit of $h\nu/m c$ per absorbed photon. Even though the photon absorption is quantized and the velocity of the atom is well measured, it is not possible to give a precise meaning to the local momentum of the photon. For a photon, like any other particle, the momentum is defined only for a plane wave. Any other field can be decomposed as a sum of plane waves. An atom placed in this field perceives all those photons, however the transferred recoil is not the average momentum of those photons. This leads to the following result: a superposition of photons each with a momentum equal to $h\nu/c$ may result in a recoil velocity larger than $h\nu/mc$ for the absorption of one photon. To the best of our knowledge, we report the first ever observation of this effect. Finally, this work is also of  special interest for applications of BOs and atom interferometry for high precision measurements \cite{Peters2001, McGuirk2002, Fixler2007,  Bouchendira2011, Duan2014, Rosi2014, Schlippert2014, Gillot2014, Barrett2016, Asenbaum2017, Bidel2018, Parker191}

\section{Acknowledgments}
This work was supported by the Agence Nationale pour la Recherche, INAQED Project No. ANR-12-JS04-0009, the Cluster of Excellence FIRST-TF and the National Institute of Standards and Technology's Precision Measurement Grant.

\bibliography{LM15738.bib}

\end{document}